# Tentative demonstration of all-silicon photodetector: from near-infrared to mid-infrared

Jiaxin Ming[1,2], Yubing Du[3,*], Tongtong Xue[4], Yunyun Dai[1,4], and Yabin Chen[1,*]

[1]*School of Interdisciplinary Science, Beijing Institute of Technology, Beijing 100081, China*.

[2]*School of Chemistry and Chemical Engineering, Beijing Institute of Technology, Beijing 100081, China*.

[3]*Beijing Aerospace Feiteng Equipment Technology Co., Ltd., Beijing 100094, China*.

[4]*School of Physics, Beijing Institute of Technology, Beijing 100081, China*

*Corresponding authors: dyblucky@163.com; chyb0422@bit.edu.cn

## ABSTRACT

Metastable silicon phases have attracted extensive attention these years, due to their fundamentally distinct photoelectric properties compared to the conventional diamond cubic (I) counterpart. Certain metastable phases, prepared via thermal heating method, can exhibit direct bandgap characteristics, significantly enhancing their light absorbance and quantum efficiency. Herein, we tentatively demonstrate an all-silicon photodetector working from near- to mid-infrared bands through precisely selective laser annealing strategy. We systematically investigated the optical properties and optoelectronic response of III/XII mixture, IV phase, and III/XII-I homojunctions. The obtained results reveal that III/XII composite and IV phase exhibit negative and positive photoconductivity, respectively. Furthermore, the established laser heating approach facilitates us to fabricate all-silicon homostructures with tunable photoconductive properties, such as III/XII-I and IV-I junctions. These findings can expand the potential applications of metastable semiconducting materials in optoelectronics and photodetectors.

## 1. INTRODUCTION

Diamond cubic silicon (Si-I) exhibits superior photoelectric response in the visible light range, leading to its widespread application in photodetectors[1]. However, due to its intrinsic bandgap of 1.12 eV[2], Si-I-based photodetectors typically present limited sensitivity at infrared wavelengths[3], significantly limiting its further application in advanced infrared photodetectors. The pressure-temperature phase diagram of silicon reveals numerous metastable silicon structures[4], which show distinct electronic and optical properties compared to Si-I, presenting promising opportunities to develop novel silicon-based photoelectric materials. Experimentally, various metastable silicon phases, including Si-III, Si-XII, and Si-IV have been successfully synthesized[5]. Some metastable phases are even proved with direct bandgap characteristics, significantly enhancing their light absorption and emitting efficiency. For example, Si-III was initially predicted to be a semiconductor with a direct band gap of 0.43 eV[6], but later theoretical studies suggested it might be a semimetal[7]. In 2017, Zhang et al[8]. successfully synthesized bulk Si-III samples and experimentally confirmed its direct bandgap nature, although the gap energy is as narrow as 0.03 eV. In contrast, research on Si-XII and Si-IV primarily remains at theoretical level, lacking the precise experimental verifications[5]. Theoretical calculations indicate that Si-XII possesses an indirect bandgap of approximately 0.24 eV[7], while Si-IV has been predicted to have an indirect bandgap of around 0.95 eV[9]. Furthermore, several studies have demonstrated that thermal annealing can treatment induce phase transformation of Si-III into Si-IV and Si-I phases at different temperatures[10,11]. These findings indicate that the optical properties of metastable silicon materials still remain controversial in the infant stage.

Based on these, this study systematically investigated the optical and photoelectric properties of metastable silicon phases (Si-III/XII and Si-IV) within the infrared spectral region, and further explored the precisely selective laser annealing strategy to fabricate all-silicon homostructures. Specifically, we evaluated the optical absorption characteristics and photodetection performance of these metastable silicon materials. Moreover, we successfully fabricated silicon-based homostructures and assessed related photoconductivity. This study provides important experimental foundation and technical support for the future development of silicon-based infrared photodetectors.

## 2. EXPERIMENTAL METHODS

### 2.1 Laser annealing process

In the laser heating experiment, a YAG laser with a wavelength of 1064 nm was used and its power was tuned as 4 W. The actual temperature was calibrated by fitting the acquired blackbody radiation curve with Planck's law. The laser heating step was set as 3 μm, and the selective heating area was adjusted according to the sample geometry. Typically, the length and width of heating area varied between 30 and 50 μm.

### 2.2 Optical and photoelectric measurements

Optical transmission experiments were conducted using a Bruker v70 Fourier spectrometer. Raman spectra were collected in a backscattering configuration using a Horiba iHR500 spectrometer. A 532 nm laser was used as the excitation source. Metal electrodes (Cr/Au 10/80 nm) were deposited onto the samples through e-beam evaporator. Photoelectric measurements were performed using a probe station equipped with a Keithley SourceMeter instrument (Keithley 2636B).

## 3. RESULTS AND DISCUSSION

### 3.1 Preparation of metastable silicon homostructures

According to the phase diagram of silicon, Si-III/XII can transition into IV phase at ~470 K, while Si-IV further converts into Si-I at ~1050 K[4,11]. By precisely selective laser annealing strategy, we applied this developed method to fabricate all-silicon homostructures. As shown in Figure 1a, the process involves four steps:(1) Synthesizing metastable Si-III/XII via dynamic decompression technique[12]; (2) Selectively apply laser annealing on Si-III/XII to locally create a III/XII–I homostructure; (3) Performing laser annealing on Si-III/XII to induce its transformation into Si-IV; (4) Conducting localized laser annealing on Si-IV to fabricate a IV-Si homostructure.

Figures 1b and 1c display optical images of the fabricated all-silicon homostructures. As observed in Figure 1d, high-quality III/XII–I homostructures were successfully fabricated through localized high-temperature annealing of III/XII. Notably, due to the thermal diffusion effect during laser heating, the diffusion region experiences lower temperatures compared to the directly irradiated area, thereby facilitating the formation of the Si-IV phase. Figure 1f illustrates the clear phase boundary in the IV–I homostructure. Furthermore, Figures 1e and 1g present the representative Raman spectra from different regions in these homostructures, confirming their phase structures and acceptable quality.

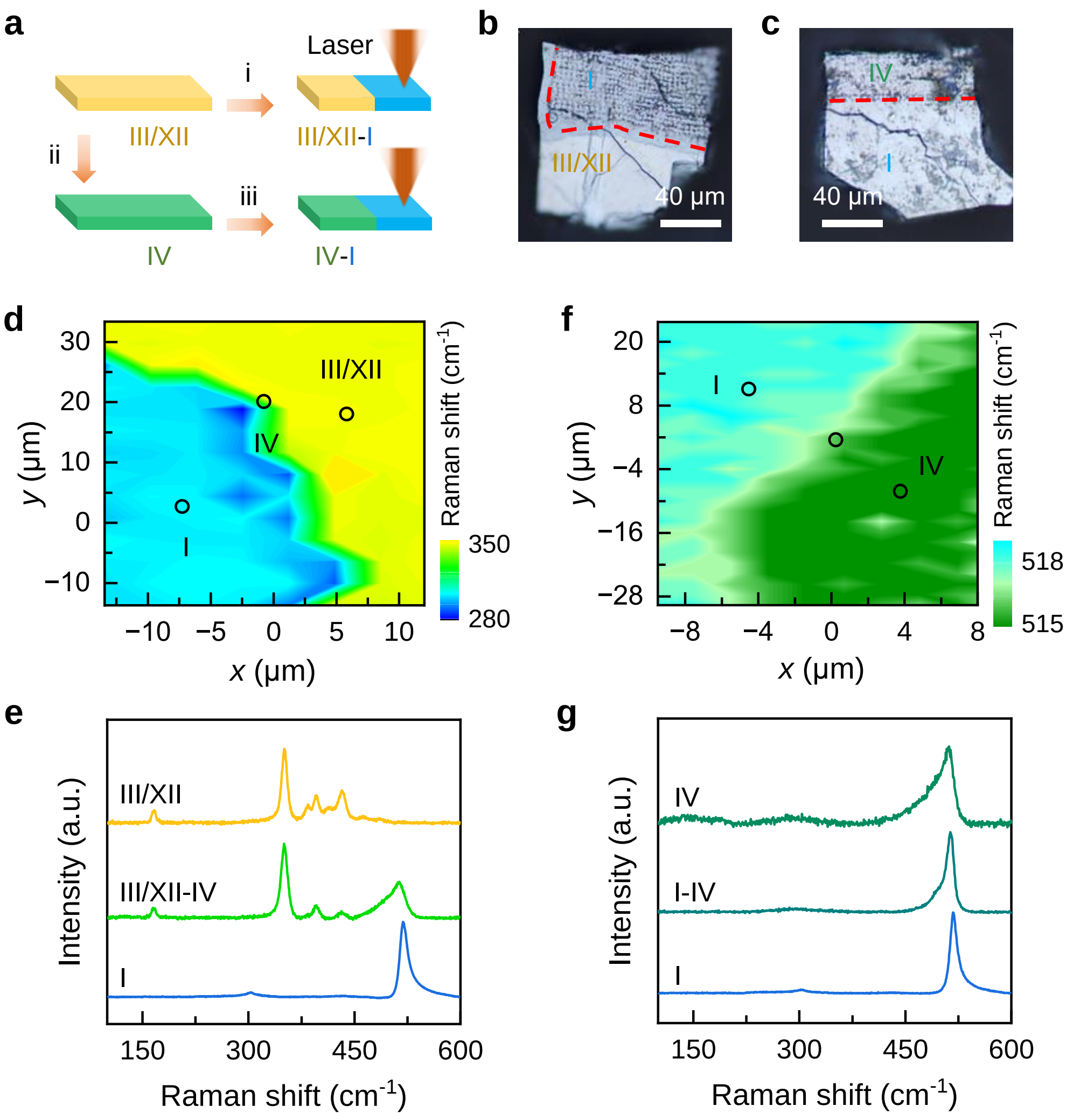

**Figure 1. Preparation of metastable silicon homostructures.** (**a**) Schematics of the preparation process of silicon homostructures via laser annealing approach. For III/XII-I homostructure, III/XII mixture (yellow) was locally heated with laser to transform to I phase (light blue), as marked by the pathway i. For the IV-I homostructure, III/XII mixture first entirely transforms to IV (green) (pathway ii), and then it is partially irradiated to form

I phase (pathway iii). (**b-c**) Representative optical image of silicon homostructure. III/XII-I (b), IV-I (c). The red dashed lines mean the phase boundaries. (**d**) Raman mapping result of III/XII-I homostructure. (**e**) Typical Raman spectrum measured from pure I, III/XII-IV interface, and III/XII area, as indicated by black circles in (d). (**f**) Raman mapping result of IV-I homostructure. (**g**) Raman spectrum measured from pure I, IV-I interface, and pure IV, as indicated by black circles in (f).

### 3.2 Photoelectric properties of Si-III/XII mixture

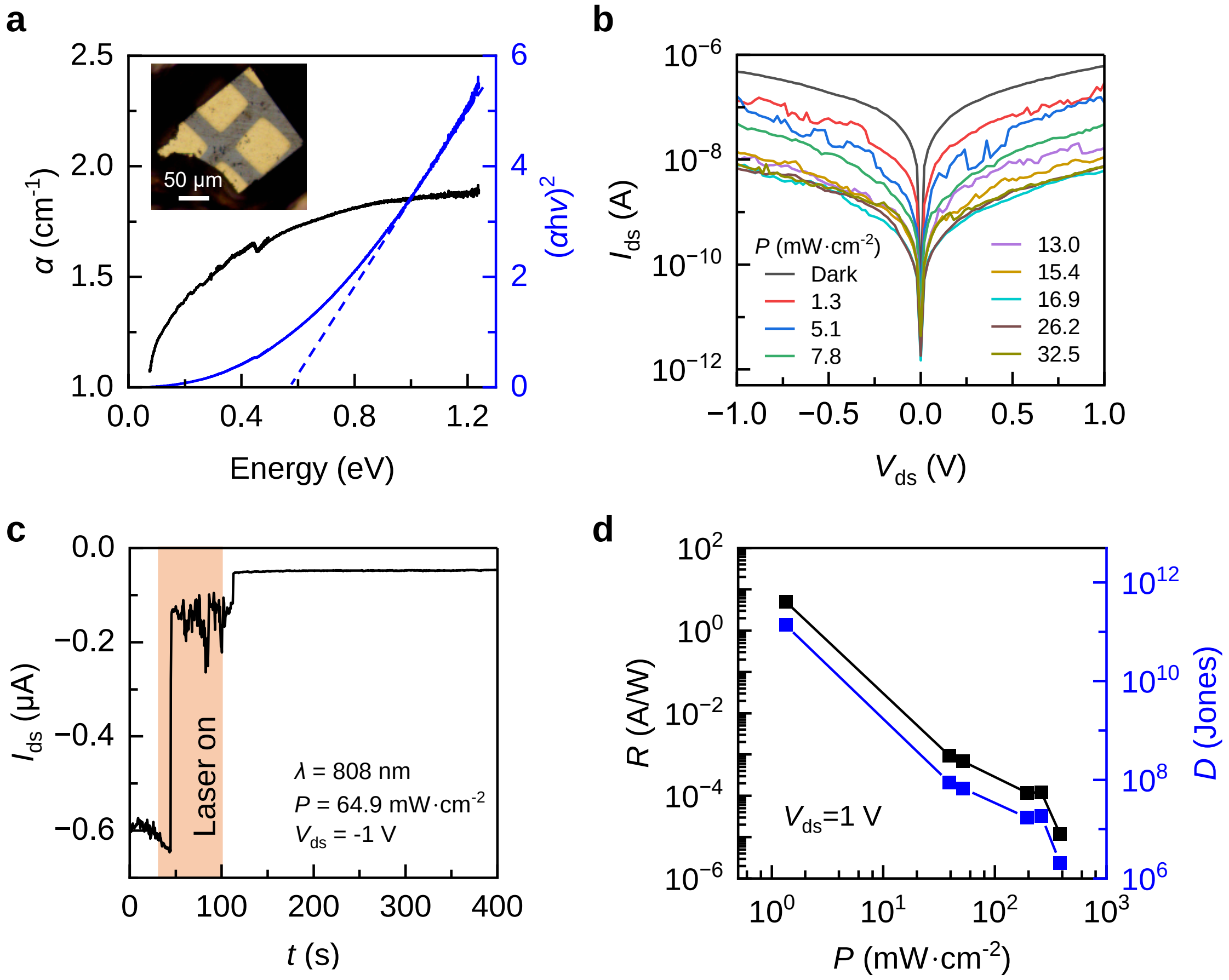

**Figure 2. Photoelectric properties of Si-III/XII mixture.** (**a**) Absorption spectra of Si-III/XII. Its direct band gap was extracted as 0.55 eV by Tauc plot method. Inset is the optical image of III/XII sample. (**b**) Output characteristic curves of Si-III/XII under different laser intensities. (**c**) Time-dependent variation of $I_{ds}$ of Si-III/XII. (**d**) Responsivity and detectivity of Si-III/XII as functions of laser intensity. The bias voltage was kept as 1.0 V.

To better understand the photoelectric properties of Si-III/XII, we conducted Fourier-transform infrared spectroscopy and photoelectric measurements. As shown in Figure 2a, the infrared transmission spectra indicate that Si-III/XII is a direct bandgap semiconductor with a bandgap of 0.55 eV, determined using the Tauc plot method. The output characteristic curves under varying laser intensities, shown in Figure 2b, reveal a negative photoconductivity effect, wherein the current decreased as the laser power rose up. Furthermore, Figure 2c illustrates that Si-III/XII cannot return to its original conductive state upon the laser was turned off, confirming the presence of persistent negative photoconductivity. In order to gain a deeper understanding of the photo-detective properties, the responsivity (*R*) and detectivity (*D*) of III/XII were calculated under different laser powers. The results indicate a maximum responsivity of 5.05 A/W and a detectivity of $1.39 \times 10^{11}$ Jones (Figure 2d). These results suggest that Si-III/XII exhibits great sensitivity and detection performance at 808 nm wavelength.

### 3.3 Photoelectric properties of Si-IV phase

Infrared transmission spectrum confirmed that Si-IV is a direct bandgap semiconductor with a bandgap of ~0.8 eV (Figure 3a). Unlike Si-III/XII, Si-IV exhibits positive photoconductivity, as shown in Figure 3b, where its photocurrent gradually increases with laser intensity and then saturates at around 103.2 $mW \cdot cm^{-2}$. Figure 3c illustrates the photoelectric response of Si-IV with a rise time of 14.4 s and a fall time of 56.4 s, significantly longer than that of crystalline cubic phase. Additionally, we calculated its responsivity and detectivity, resulting to a maximum responsivity of 0.02 A/W and a detectivity of $7.87 \times 10^{9}$ Jones at a laser power density of 0.67 $mW \cdot cm^{-2}$ (Figure 3d). Both of responsivity and detectivity significantly degrade as laser power.

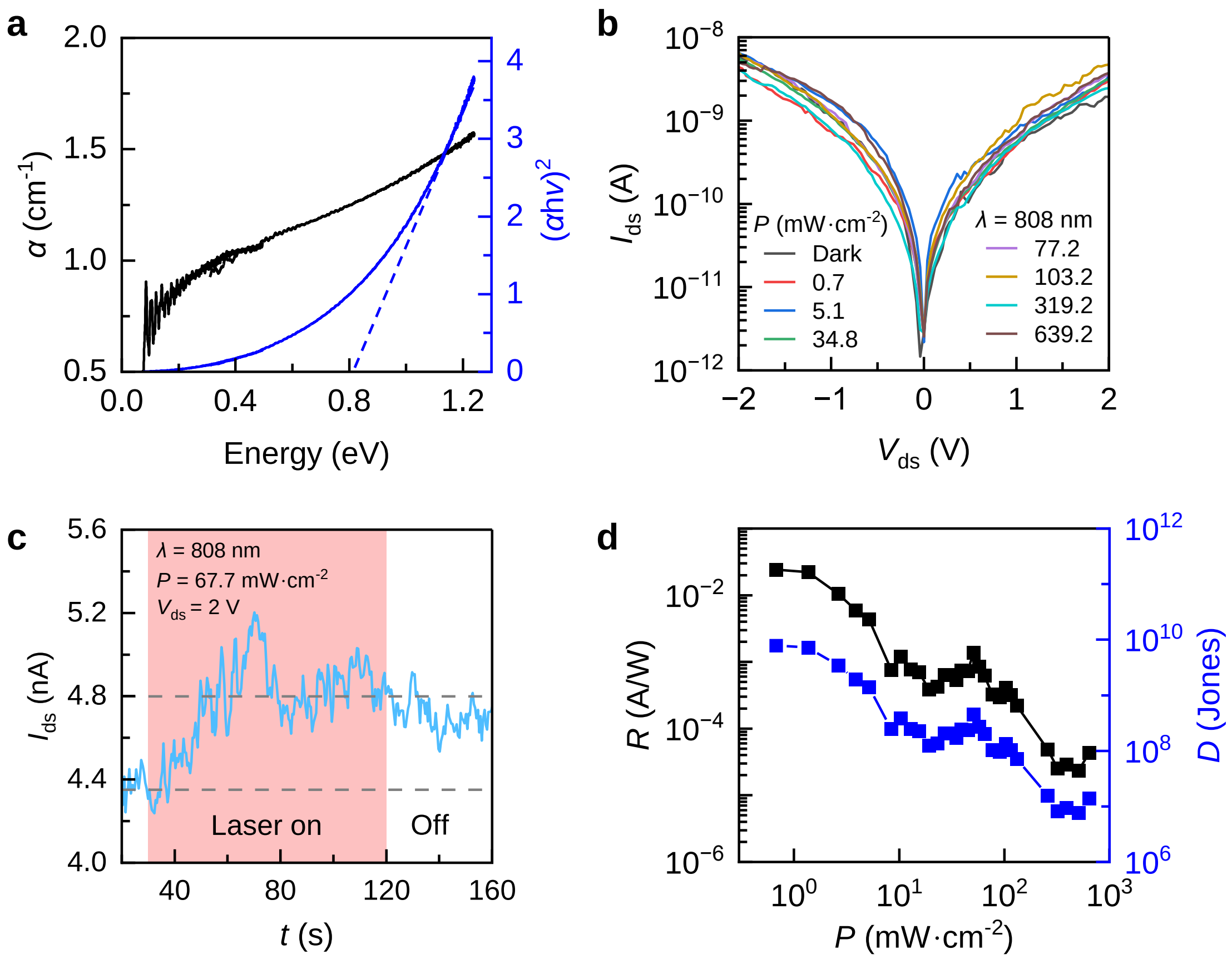

**Figure 3. Photoelectric properties of Si-IV phase.** (**a**) Absorption spectra (black) of Si-IV, with the direct bandgap of 0.79 eV extracted using the Tauc plot method (blue). (**b**) Output characteristic curves of Si-IV at different laser intensities. (**c**) Time-dependent variation of $I_{ds}$ in Si-IV under 808 nm laser. (**d**) Responsivity (*R*) and detectivity (*D*) of Si-IV as a function of incident laser intensity.

### 3.4 Photoelectric properties of Si-III/XII-Si-I homostructure

The photoelectric response of the Si-III/XII–Si-I homostructure was investigated under varying laser intensities, as shown in Figure 4. Within the 0–1 V range, the current remained nearly suppressed; while in the 1–2 V range, it exponentially increased with voltage (Figure 4b), indicating the rectification effect. This behavior is likely attributed to the complex carrier transport process, where charge carriers may traverse multiple nanocrystalline boundaries[13]. Moreover, the current increased with laser intensity, confirming the positive photoconductivity of this homostructure (Figure 4c).

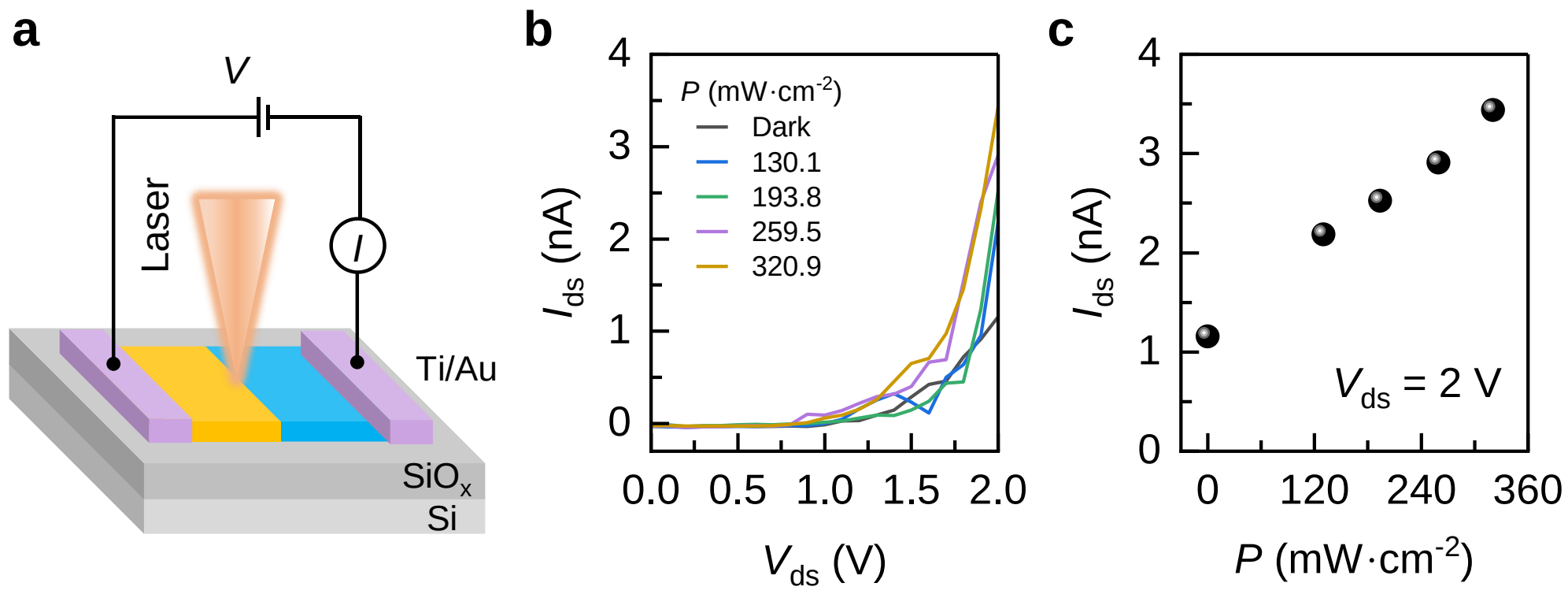

**Figure 4. Photoelectric properties of silicon III/XII-I homostructure.** (**a**) Schematic diagram of the photoelectric testing device. (**b**) Output characteristic curves of the III/XII-I homostructure with different laser intensities. (**c**) Variation of $I_{ds}$ of the III/XII-I homostructure as a function of laser power density at $V_{ds}$ = 2 V.

### 3.5 Comparison of photoresponsivity

A detailed comparison of the responsivity and detectivity of III/XII mixture and IV phase is shown in Figure 5, referred with the annealed I phase (573 K for 2 h under 7 GPa). It is obvious that III/XII mixture exhibits superior photoelectric performance than IV phase at 808 nm laser, primarily attributed to its smaller optical bandgap. Among these materials with positive photoconductivity, the annealed Si-I phase demonstrates inferior responsivity and detectivity performance, compared to the metastable IV phase.

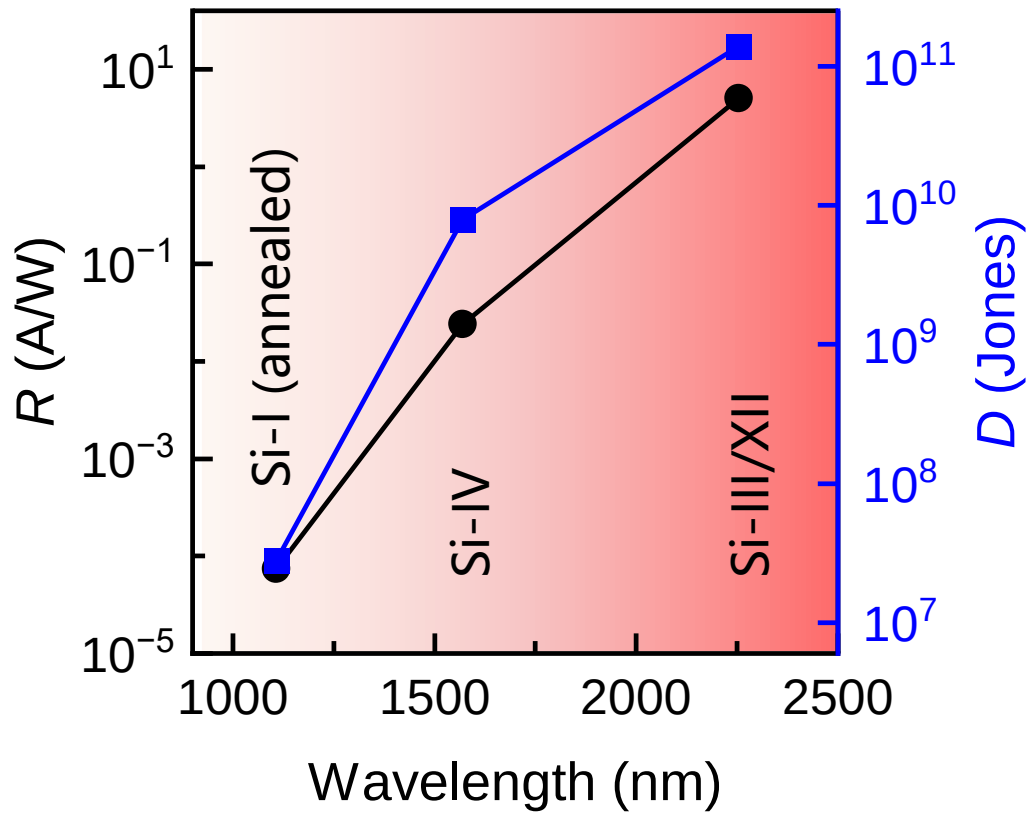

**Figure 5. Comparison of photoresponsivity.** Comparison of III/XII mixture, IV phase and annealed I phase in responsivity and normalized detectivity. Si-I was treated at 7 GPa and 573 K for 2 h at high temperature and high pressure.

## 4. CONCLUSION

In conclusion, we successfully fabricated all-silicon homostructures by precisely controlling annealing conditions, and systematically investigated the photodetection performance of III/XII mixture, IV and their homojunctions. Our findings reveal that III/XII mixture and IV are direct bandgap semiconductors with bandgaps of 0.55 and 0.79 eV, respectively. Si-III/XII exhibited negative photoconductivity and exceptional photoelectric performance with its detectivity up to ~$1.4\times10^{11}$ Jones. Furthermore, as a positive photoconductive material, Si-IV exhibits superior responsivity and detectivity compared to the annealed Si-I. This study can provide new opportunities for advancing silicon-based photodetector technologies and expanding their applications in infrared optoelectronics.

## ASSOCIATED CONTENTS

### Supporting Information

The Supporting Information is available free of charge online.

### Acknowledgements

This work was financially supported by the National Natural Science Foundation of China (grant numbers 52472040 and 52072032). The authors acknowledge the helpful discussion and technical support from Dr. K. Zhang and Dr. W.D. Hu (Shanghai Institute of Technical Physics, China).

### Notes

The authors declare no competing financial interest.

### Data Availability

All data related to this study are available from the corresponding author upon reasonable request.